\documentclass{article}

\usepackage{arxiv}

\usepackage[utf8]{inputenc} 
\usepackage[T1]{fontenc}    
\usepackage{hyperref}       
\usepackage{url}            
\usepackage{booktabs}       
\usepackage{amsfonts}       
\usepackage{nicefrac}       
\usepackage{microtype}      
\usepackage{lipsum}
\usepackage{graphicx}
\graphicspath{ {./images/} }

\usepackage{cite}
\usepackage{amsmath,amssymb,amsfonts}
\usepackage{algorithmic}
\usepackage{subfig}
\usepackage{textcomp}
\usepackage{xcolor}

\title{Preliminary Study of Connectivity for Quantum Key Distribution Network}

\author{
 Ying Liu \\
  Computational Science Initiative\\
  Brookhaven National Laboratory\\
  Upton, NY 11973 \\
  \texttt{yingl@rutgers.edu} \\
}

\begin{document}
\maketitle
\begin{abstract}
Quantum network is fragile to disturbances when qubits are transmitted through quantum channel. Reliability is an essential requirement for a quantum network and even the future quantum internet. A metric is needed to describe the reliability of a quantum network to build a robust infrastructure and communication protocols. In this work, we combined quantum physical parameters with graphic algebraic connectivity to indicate the transmission throughput of a grid quantum network. This metric can be extended to multiple equal point-to-point distance topology. This work also studies how to tune specific physical parameters to maintain or even increase connectivity when nodes or edges are removed from a network. Using this metric, resources consumption are compared.   
\end{abstract}


\section{Introduction}
Due to the large de-coherence property of a single photon over long distances, the development of quantum network lags quite behind quantum computer. The quick decay of correlation beyond coherent time caused by outside-world interference makes the reliability, one of the key issues for achieving a practical quantum network. Quantum Key Distribution (QKD) is the first quantum network protocol designed for quantum communication, which has been studied since 1980s. 

This work aims to understanding the secure connectivity of a quantum network when QKD protocol is operating in its physical layer. QKD includes various key exchange protocols to establish a shared key before secure communication is taken between two direct connected parties. Although only one qubit can be transmitted each time, it is supposed that, after repeated transmission, a usable length of key / information can be shared between two parties. It is obvious that the number of repeats until a specified length of information is achieved, indirectly relates to connectivity. 

There are several factors that cause errors when a qubit transmitted to the end: 1) non-clone property of quantum, that is, as long as a quantum wave is measured, the information carried by a qubit on-the-fly can be distorted or evenly destroyed. A known example is eavesdropping. This behavior causes a link or node removal, depending on the target of eavesdropper, in a quantum network; 2) resonance caused by other concurrent transmitted wave functions; 3) quantum transmitter such as a laser, produces inaccurate or distorted quantum pulse; 4) quantum detector which has a low receiving fidelity, such as optical imperfections, and dark counts. There are already few papers discussing it \cite{bagraev2005parameter, gisin2002quantum, aguiar2013network, gobby2004quantum}. If qubits are transmitted through free space, more physical error sources need to take into consideration, such as signal strength deterioration caused by a quantum channel, and environmental noise.  

Quantum channel can either be entangled or not entangled. Regarding entangled channel, before transmitting information, an entangled pair of photons are shared between transmitter and receiver. Next, an entangled qubit carrying information is transmitted by associating with the entangled channel. After receiving the new entangled pair, corresponding numerical operations are performed to obtain the original sent information, which is the so-called quantum teleportation.

Although entangled quantum teleportation can reduce the re-transmission, its real-world implementation over long-distance still needs great efforts. Current implemented quantum network protocols are through non-entangled channel, especially QKD. Most proposed network protocols which integrate quantum security \cite{elliott2018darpa} are based on BB84 \cite{shor2000simple}.  

\section{Background: quantum bit rate and channel capacity}

\subsection{Secrete key rate based on BB84}
Quantum bit error rate (QBER) is introduced by \cite{gisin2002quantum} to compare different QKD setups. For a non-entangled channel based on BB84 protocol, the two-point QBER using faint laser pulses is according to \cite{gisin2002quantum}

\begin{equation}
    QBER = \frac{R_{err}}{R_{sift} + R_{err}} \approx \frac{R_{err}}{R_{sift}}
\end{equation}

where $R_{sift}$ is the sifted key rate which is counted when the Alice (A) and Bob (B) choose the same basis. Thus, in BB84 protocol, the $R_{sift} = R_{raw} / 2$. $R_{raw}$ is the raw data rate which equals to the number of photons per unit time multiplying the successful probabilities from sending photons to correctly receiving it. Thus, $R_{raw} = f_{ref} \mu t_{link}\eta$ \cite{gisin2002quantum}. $f_{ref}$ is the frequency of a laser to produce a pulse. $\mu$ is the average number of photon carried by the pulse. $t_{link}$ is the successful probability that photons arrive at the detector. $t_{link}$ relates to the channel conditions (i.e. the coherence decreases exponentially with distance \cite{gobby2004quantum}) and eavesdropper's measurement. $\eta$ is the probability that photons' are detected. Thus,     

\begin{equation}
    R_{sift} = \frac{1}{2} f_{ref} \mu t_{link}\eta  
\end{equation}
where sending rate $R_{s}$ is $f_{ref} \mu$. 

For non-entangled photon stream,
\begin{equation}
    R_{err} = R_{opt} + R_{det}
\end{equation}
where $R_{opt}$ is the error rate that photons fall into the wrong detector due to optical imperfections. Assume $p_{opt}$ to be the probability that the wrong detector captures photon. Therefore,

\begin{equation}
    R_{opt} = R_{sift}p_{opt}
\end{equation}

Based on \cite{gisin2002quantum}, $R_{det}$, which is the error rate caused by dark count, equals to

\begin{equation}
    R_{det} = \frac{1}{4} f_{ref} p_{dark} n
\end{equation}
where $p_{dark}$ is the probability of valid dark count per detector. $n$ is the number of detectors at a receiver. Thus,  

\begin{equation}
    QBER = p_{opt} + \frac{1}{2} \frac{p_{dark}n}{\mu t_{link}\eta }
\end{equation}

Based on \cite{scarani2009security} and \cite{vedovato2015space}, the secret key rate $R$ equals to

\begin{equation} \label{eq:R}
    R = 1 - 2h(QBER)
\end{equation}
where $h(x) = -x log_{2}x - (1-x) log_{2} (1-x)$ is the binary entropy. 

\subsection{Quantum Channel Capacity}
A channel (optical / free space) in a quantum system can either be with or without pre-shared entangled pairs. The capacity under these two channels are defined slightly differently based on the input states and formats. 

When the input is in product states (no entanglement), and the outputs of the channels (including multiple parallel channels) are measured jointly, the channel capacity according to \cite{gyongyosi2018survey} is as follows:  

\begin{align} \label{eq:C_N}
    C(N) & = \max_{p_{i},\rho_{i}}\chi \\ 
         & = \max_{p_{i},\rho_{i}} [S(N(\sum_{i}p_{i}\rho_{i})) - \sum_{i}p_{i}S(N(\rho_{i}))] \\
    subject \; to \; \\ 
         & \sum_{i}p_{i} = 1
\end{align}
where $N(.)$ is the channel function, which describes how quantum information interacts with environment. For example, it can be either unitary transformation or other distortions. $S(.)$ is the von Neumann entropy. $\rho = \sum_{i}p_{i}\rho_{i}$ is the density matrix of input quantum states. A quantum communication system should be designed to obtain the maximum channel capacity over all possible combinations of basic ensembles. 

For a channel with pre-shared entangled pairs, the quantum channel capacity does not need asymptotic version. Thus, $C_{E}(N)$, which equals to $\lim_{n\to\infty} \frac{1}{n} \chi (N^{\bigotimes n})$, referencing from \cite{gyongyosi2018survey} and \cite{shor2003capacities}, becomes  
\begin{align} \label{eq:C_E}
    C_{E} (N) & = C_{E}^{(1)}(N) = \max_{p, \rho_{i}} I(A:B) \\
              & = \max_{\rho \epsilon H_{in}} S(\rho) + S((N(\rho))) - S((N \otimes I)(\Phi _{\rho}))
\end{align}
$I(A:B)$ is the maximum mutual information between Alice and Bob. Mutual information measures how much B can tell about A. $C_{E}^{1}(N)$ means capacity of one channel. Here (\ref{eq:C_N}) and (\ref{eq:C_E}) only describe the one-hop communication capacity. 

\subsection{Relations between R and C}
Secrete key rate described by (\ref{eq:R}) only considers sending and receiving rate of qubits under the assumption of perfect channel conditions, which have no channel distortions, decays and errors. Intuitively, this is an ideal case. When photons are transmitted in optical fiber and not vacuum space, the impact of channel on signals must be considered.   

Shannon theory describes the relationship between transmission rate and channel capacities to achieve small error during information transmission. Therefore, C is used as a guidance to design error-free channel codes. Thus, the relationship is as followings 
\begin{equation} \label{eq:shannon}
    \left\{\begin{matrix} exist \;errors \; & R \geqslant C \\ 
 no \; or \; exponentially \; small \; errors & R <C
\end{matrix}\right.
\end{equation}

(\ref{eq:shannon}) can be used to indicate the quality of a two point communications. If $R >= C$ appears frequently within $T$ time period, the link quality or weight is statistically poor in average. Since quantum channel is fragile and stochastically not stable, we use the following metric to describe the average transmission ability of a link over $T$.  

\begin{equation} 
   w_{t} = \frac{ \left \{ R_s < C \right \}_{t}}{ \left \{ R_s < C \right \}_{t} +  \left \{ R_s >= C \right \}_{t}}
\end{equation}

\begin{equation} \label{eq:link_weight}
    t_{link} = W_{t + 1}  = \alpha \cdot w_{t} + (1 - \alpha) W_{t}
\end{equation}

\begin{equation}
    w_{0}  = \left\{\begin{matrix}
1  &   R_{s0} < C_{0}\\ 
0 &   R_{s0}  \geqslant C_{0}
\end{matrix}\right.
\end{equation}
where $\alpha$ is the learning rate, which should be less than one. 

\section{Quantum network connectivity}
\subsection{Background: graph connectivity}
The (\ref{eq:C_N}) and (\ref{eq:C_E}) only quantify the point-to-point channel capacity. Another interesting topic is the capacity of a multi-hop network, which considers the end to end network transmission ability and connectivity. 

Network connectivity is usually studied through graph theory. For a static network, its connectivity is defined by algebraic connectivity / Fiedler value, which is  the second smallest eigenvalue of the Laplacian matrix, $L(V,E)$, of a network's topological graph, $G(V,E)$ where $V$ is the graph vertex set and $E$ is the edge set for $V$. 

The Fiedler value \cite{liu2016maintaining} is always non-negative, and its value is zero if and only if the graph is disconnected, in which case the number of zero eigenvalues of $L$ equals the number of connected components in a graph (e.g. a disconnected network consisting of two subnetworks that are themselves topologically connected, would yield have two eigenvalues equal to zero). Referring to \cite{fiedler1973algebraic}, the Fiedler value, represented by $\lambda^{1}$, of a graph, $G$ can be obtained by the following eigenvalue optimization problem.
\begin{equation} \label{eq:fiedler}
\begin{aligned}
& \lambda_{1} =  \min y^{T}L(V,E)y \\
& st. \; y^{T}y = 1\;\;{and}\;\;y^{T}\textbf{1} = 0
\end{aligned}
\end{equation}
where $y$ is a vector which does not equal to $\textbf{1}$.

The Laplacian matrix of a given graph is defined as follows:
Given a graph $G(V,E)$ without self cycles and multiple links between two nodes, the Laplacian matrix $L$ is calculated by
\begin{equation} \label{eq:Laplacian}
L(V, E)= D(V, E) - A(V, E)
\end{equation}

\noindent where $D(V, E)$ is a diagonal matrix whose diagonal entry contains the degrees for each node. $A(V,E)$ is the adjacency matrix with each entry being a value of zero or one when nodes are connected to each other. In addition, its diagonal is zero since, for a communication network we assume that $G(V, E)$ has no self cycles. A valuable characteristic of Laplacian matrix is that it is semi-definite. Thus, (\ref{eq:fiedler}) is solvable. 

\subsection{Quantum network connectivity in physical layer}
We use algebraic connectivity, in terms of R and C, to study the capacity connectivity of a quantum network. We formulate the problems as follows: 

\begin{equation} \label{eq:fiedler_phy}
\begin{aligned}
& \lambda_{1}^{t} =  \min y^{T}L(V (R^{t}) , E (C^{t}))y \\
& st. \; y^{T}y = 1\;\;{and}\;\;y^{T}\textbf{1} = 0
\end{aligned}
\end{equation}

$V(R^{t})$ is the vertex set of a quantum network, which can be senders (say, lasers, repeaters) and receivers / detectors. $E(C^{t})$ which is the edge set, indicating communication rate / quality of a link. 

Thus, we define each entry of adjacency matrix by (\ref{eq:R}) and (\ref{eq:link_weight}). $L$ is a weighted matrix, whose weights are the function of transmission rate, $R^{t}$, and channel capacity, $C^{t}$, which reflects the link quality.

\subsection{Quantum network connectivity resilience}
Quantum network is fragile to eavesdroppers, environment disturbances, and decoherence over long time and distance transmission, no matter the channel is entangled and not. The eavesdropper's observation can immediately destroy transmitted wave function. These failures, from network point of view, equals to removing links and edges from a graph, which has direct effects on network connectivity, and information transmission.   

Resilient connectivity of a network means the information transmitted from A can still be received at B after several links and nodes fail on the original path. However, normally it's time consuming to exam all redundant paths among two nodes in a network. Thus, an approximate criterion is to ensure the network connectivity does not drop or drop less after links and nodes failed in a network. 

Algebraic connectivity provides good properties and measuring criterion to indicate connectivity. The larger value of algebraic connectivity indicates stronger network resilience. 
The properties of algebraic connectivity that relate to the comparison of a network connectivity, are: 1) If the network's links are broken because of disturbances, the Fiedler value $\lambda_{1}(V, E_{1}) \leqslant \lambda_{1}(V, E)$, where $E_{1} \subseteq E$. This also informs that the Fiedler value can become larger when adding edges to a graph; 2) Fiedler value's upper bound is limited to the minimum degree of nodes and the total number of nodes existing in a network. Thus, it also indicates that increasing the degrees for all nodes is ineffective since the upper bound is constrained by the minimum value of the nodes' degrees; 3) the algebraic connectivity of the remaining graph after nodes are removed, is comparable as long as the number of remaining nodes in a network is same.

After identifying the weakest connectivity node, which means removing it would have the most harmful impact on the network's algebraic connectivity, we can add classical router around the node to strengthen the quantum network connectivity. 

\section{Numerical evaluation}
\begin{figure} 
\centering
\includegraphics[width=0.8\textwidth]{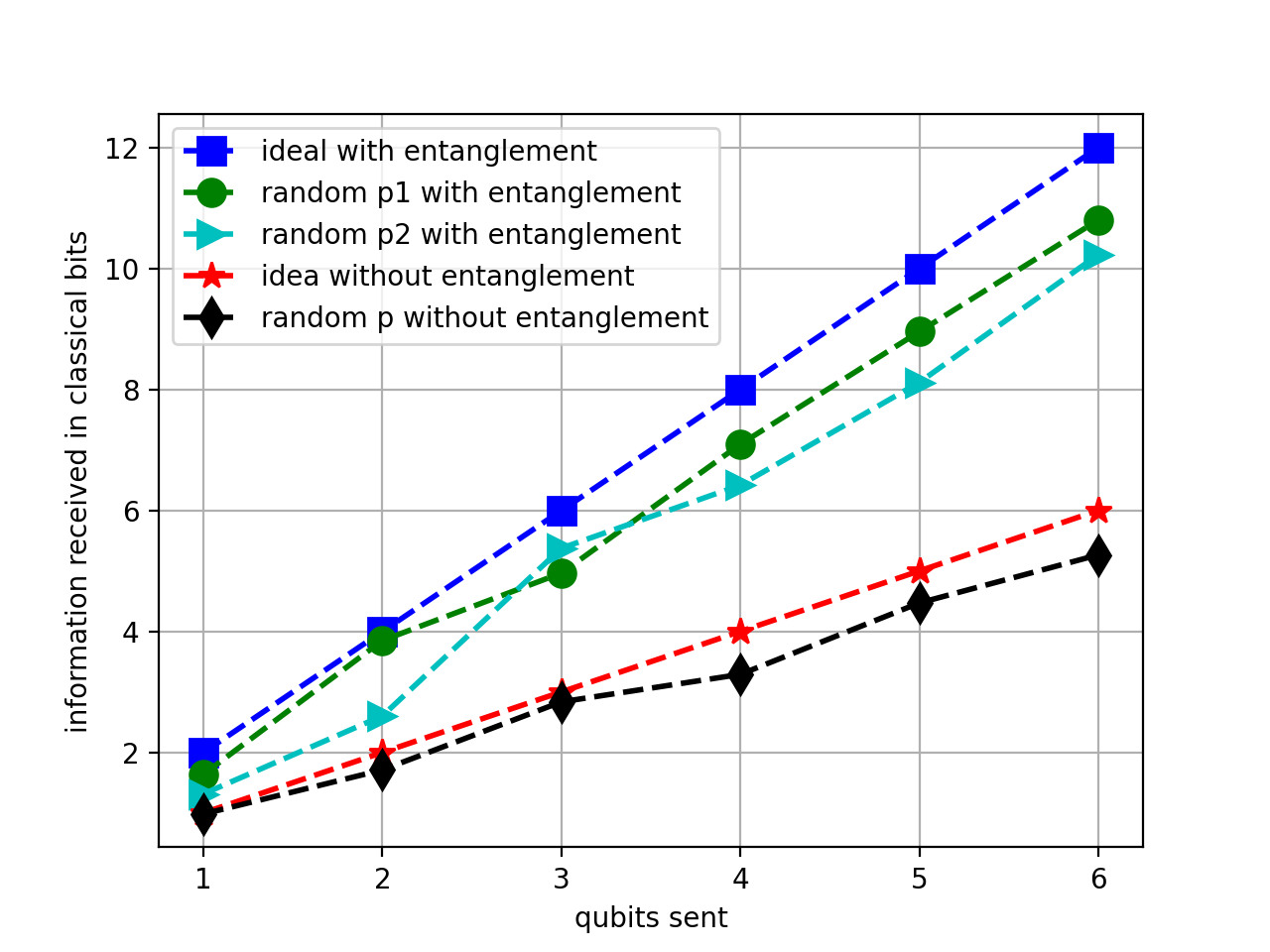}
\caption{quantum link capacity} \label{fig:qlink}
\end{figure}

We have simulated both (\ref{eq:C_N}) and (\ref{eq:C_E}) shown in Fig. \ref{fig:qlink} for two-point link capacity. In both with and without entanglement, unitary channel matrix is applied, such as identity matrix, $I$, or no-loss rotation matrix, $U$. In the two ideal cases (blue, red), we optimized $p_{i}$ for all ensembles of bases to find the maximum capacities. We derive that the entangled capacity is factor of 2 of sent qubits, which suggests the advantage of pre-distributing entangled pairs before real information is communicated. If randomly pick $p_{i}$, it indicates that specific $\mathbf{p}$ favors specific qubit input.        

\begin{figure}[h] 
\centering
\includegraphics[width=0.78\textwidth]{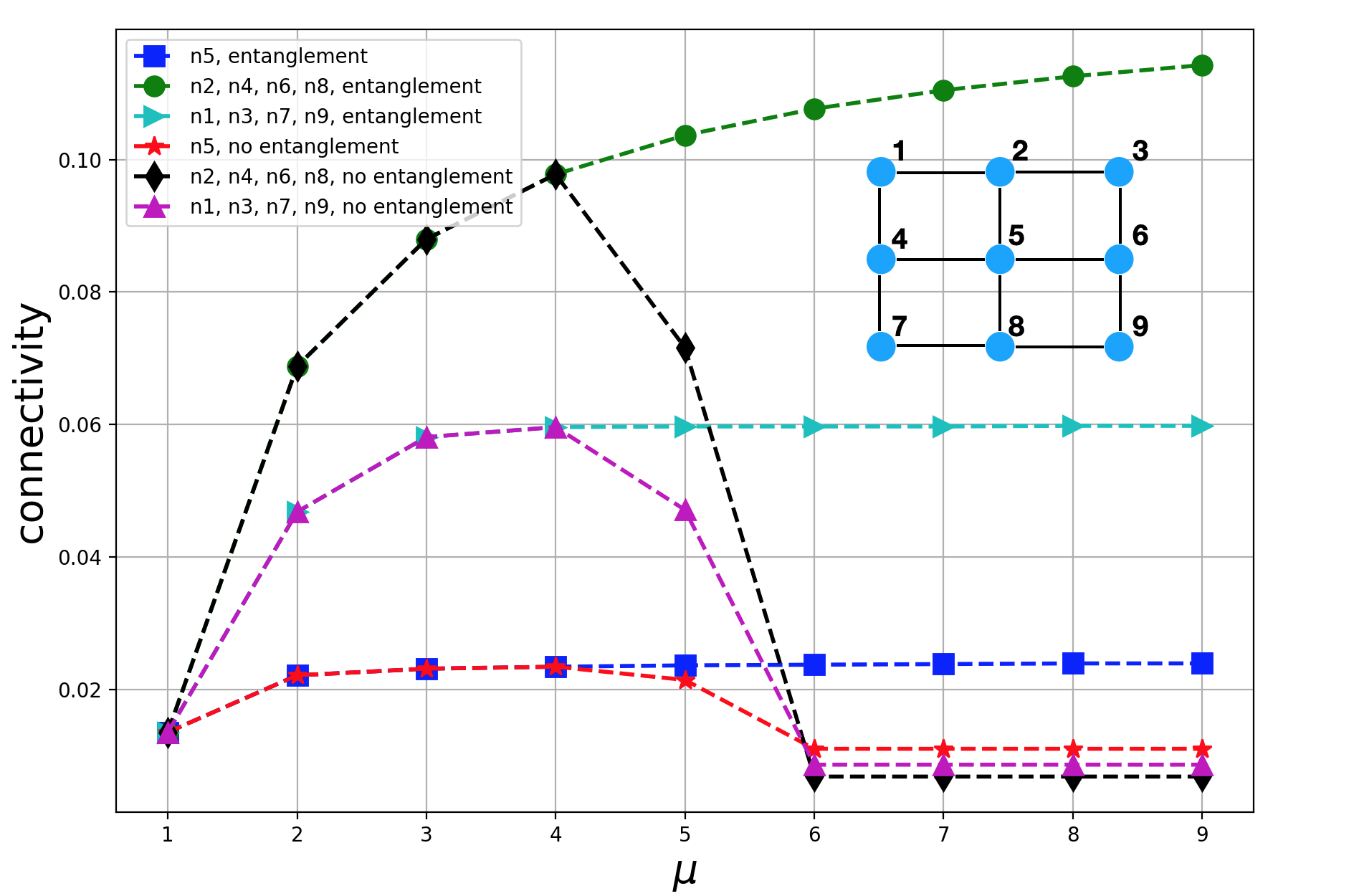}
\caption{increasing connectivity over $\mu$} \label{fig:c_incr}
\end{figure}

We studied the grid topology with 3 by 3 nodes. Node is numbered in the order of [[1, 2, 3], [4, 5, 6], [7, 8 ,9]], from left to right, and top to bottom. Fig. \ref{fig:c_incr} shows we increase $\mu$ of [5], [2, 4, 6, 8], [1, 3, 7, 9] respectively with entangled and non-entangled link capacities. We group them according to their distance (hops) from the center of the topology. Capacity is learned based on (\ref{eq:link_weight}) by 100 times with learning rate 0.5. All $\mu$ of other nodes are 1. We have obtained: 1) with increase of $\mu$ or any parameters that can increase $R_s$, [2, 4, 6, 8], one hop neighbors, can improve the connectivity most for the grid topology, even if connectivity of other group does not increase after $\mu > 4$; 2) although we increase more than 3 nodes of [2, 4, 6, 8] comparing to only improving that of one node, 5, the improvement of connectivity is more than three times. We think the reason is that the average closeness of a node to other nodes plays a role; 3) For non-entanglement case, connectivity drops dramatically after certain threshold, even if $\mu$ increases. The reason is the average learned capacity, $C_N$ become less than $R_s$ with $\mu$ increases. We expect the same phenomenon would occur for entangled channel with $\mu$ increases more. In conclusion, we believethat  properly selecting nodes could improve connectivity while the same resources, $\mu$ are used. 

\begin{figure}[h] 
\centering
\includegraphics[width=0.8\textwidth]{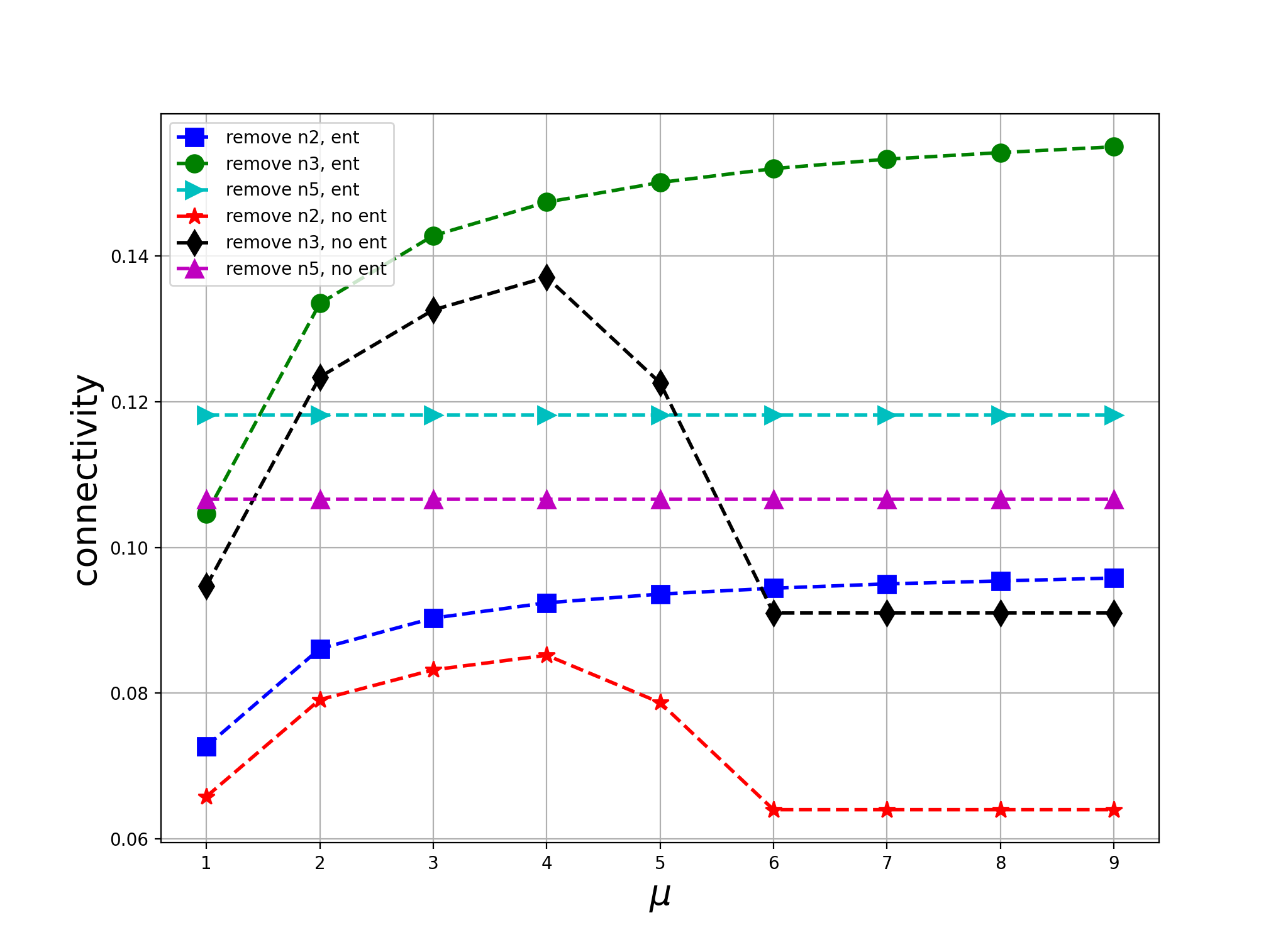}
\caption{increase connectivity by node [5] after removing [2], [3], [5]} \label{fig:rm_5}
\end{figure}

\begin{figure}[h] 
\centering
\includegraphics[width=0.8\textwidth]{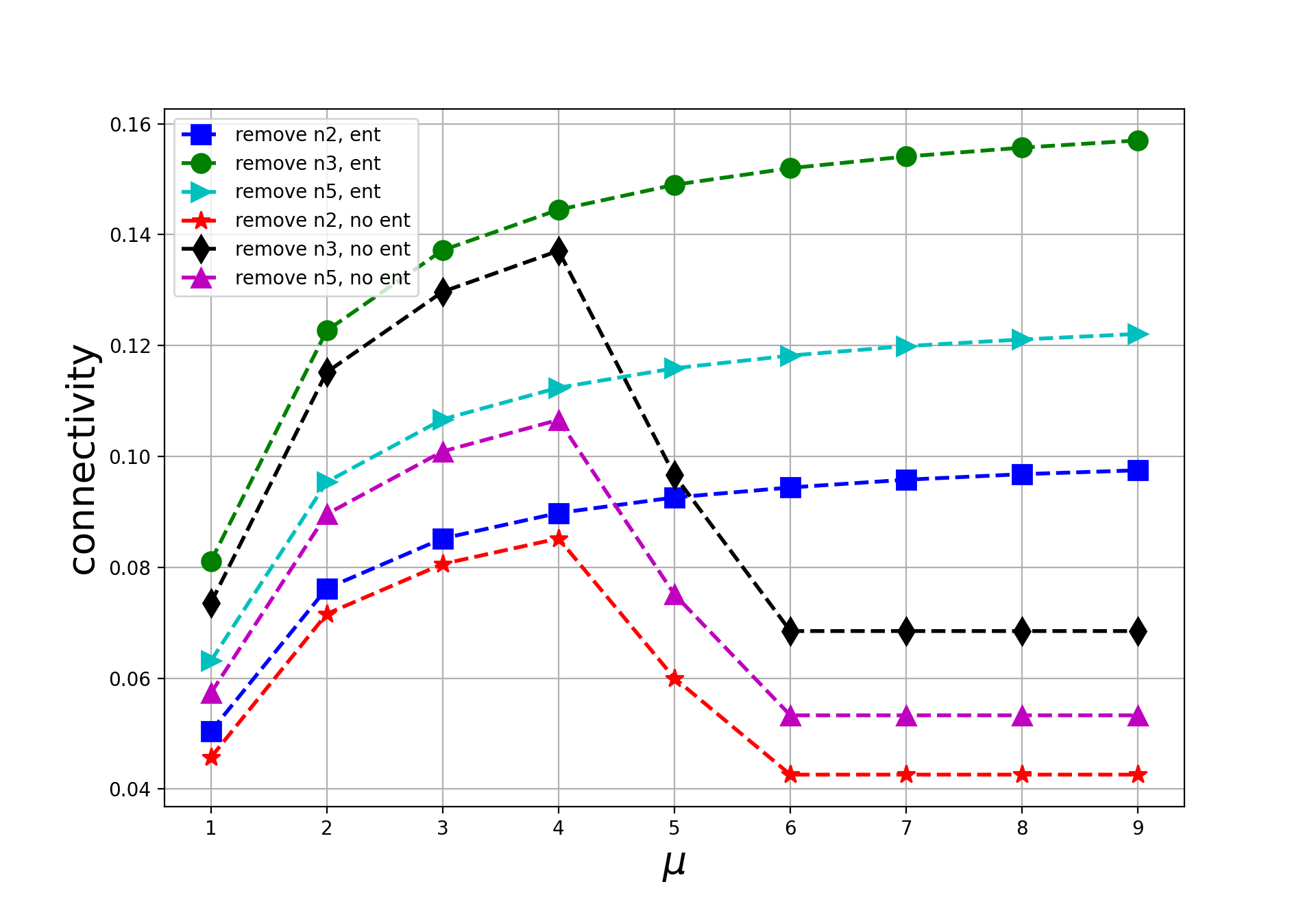}
\caption{increase connectivity at nodes [2,4,6,8] after removing [2], [3], [5]} \label{fig:rm_2468}
\end{figure}

\begin{figure}[h] 
\centering
\includegraphics[width=0.8\textwidth]{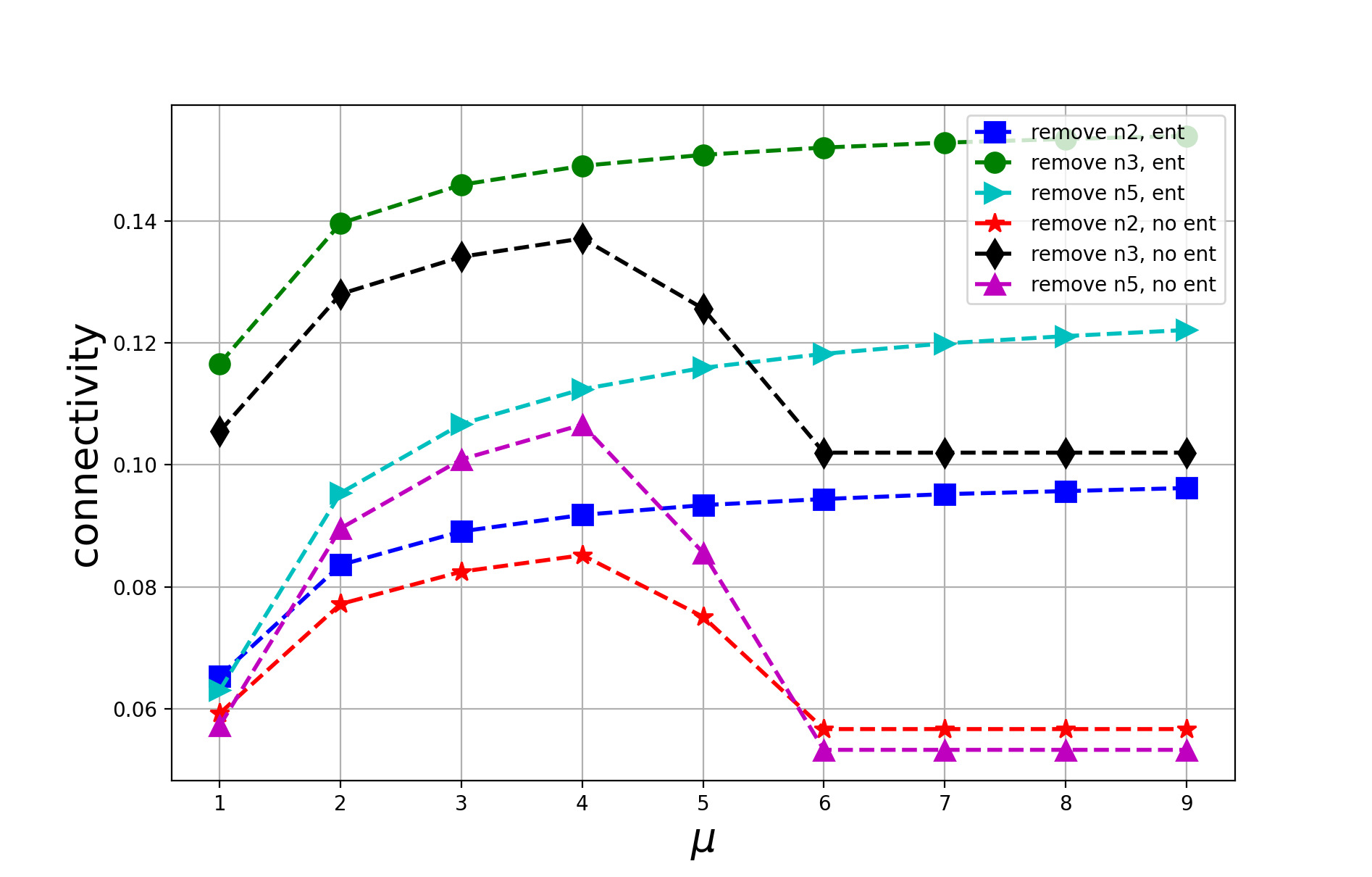}
\caption{increase connectivity at nodes [1,3,7,9] after removing [2], [3], [5]} \label{fig:rm_1379}
\end{figure}

\begin{figure}[h] 
\centering
\includegraphics[width=0.8\textwidth]{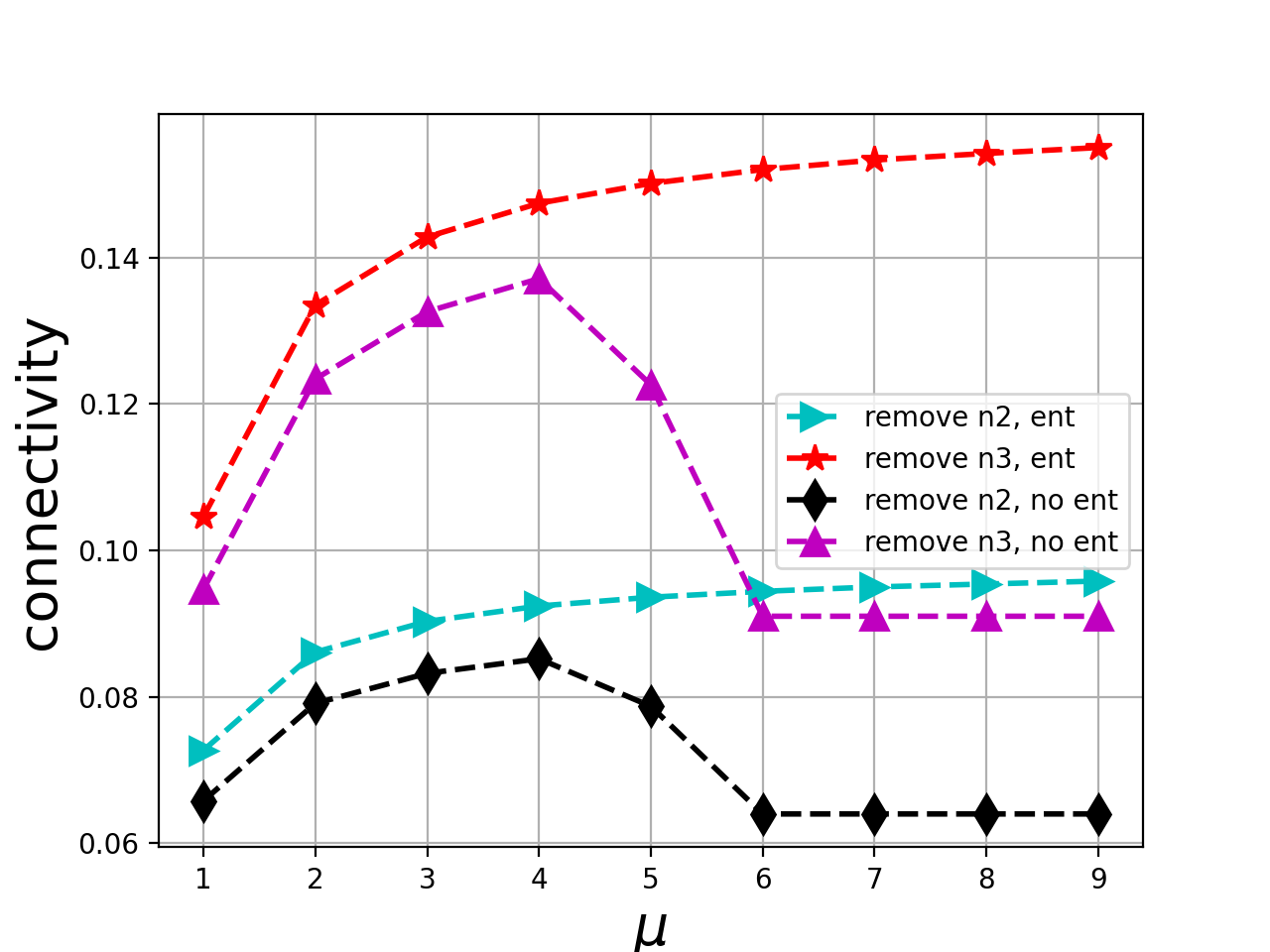}
\caption{Compare harmfulness of removing nodes on connectivity} \label{fig:harmfulness}
\end{figure}

Fig. \ref{fig:rm_5}, \ref{fig:rm_2468} and \ref{fig:rm_1379} shows how connectivity varies after removing one node from the network while increasing $\mu$ of group [5], [2, 4, 6, 8] and [1, 3, 7, 9]. Fig. \ref{fig:rm_5} shows the harmful impact of removing node 2 is larger than that of node 3. It even brings the connectivity with entangled channel below that of non-entangled one, which is not a usual case. Increasing $\mu$ of central node [5] has made small difference in terms of connectivity. From the same starting $\mu = 1$, the order of harmfulness of removing one node is [2] $>$ [3] $>$ [5]. 
Fig. \ref{fig:harmfulness} proves again that removing node 2 harms the connectivity more than node 3.  

\begin{figure}[h] 
\centering
\includegraphics[width=0.9\textwidth]{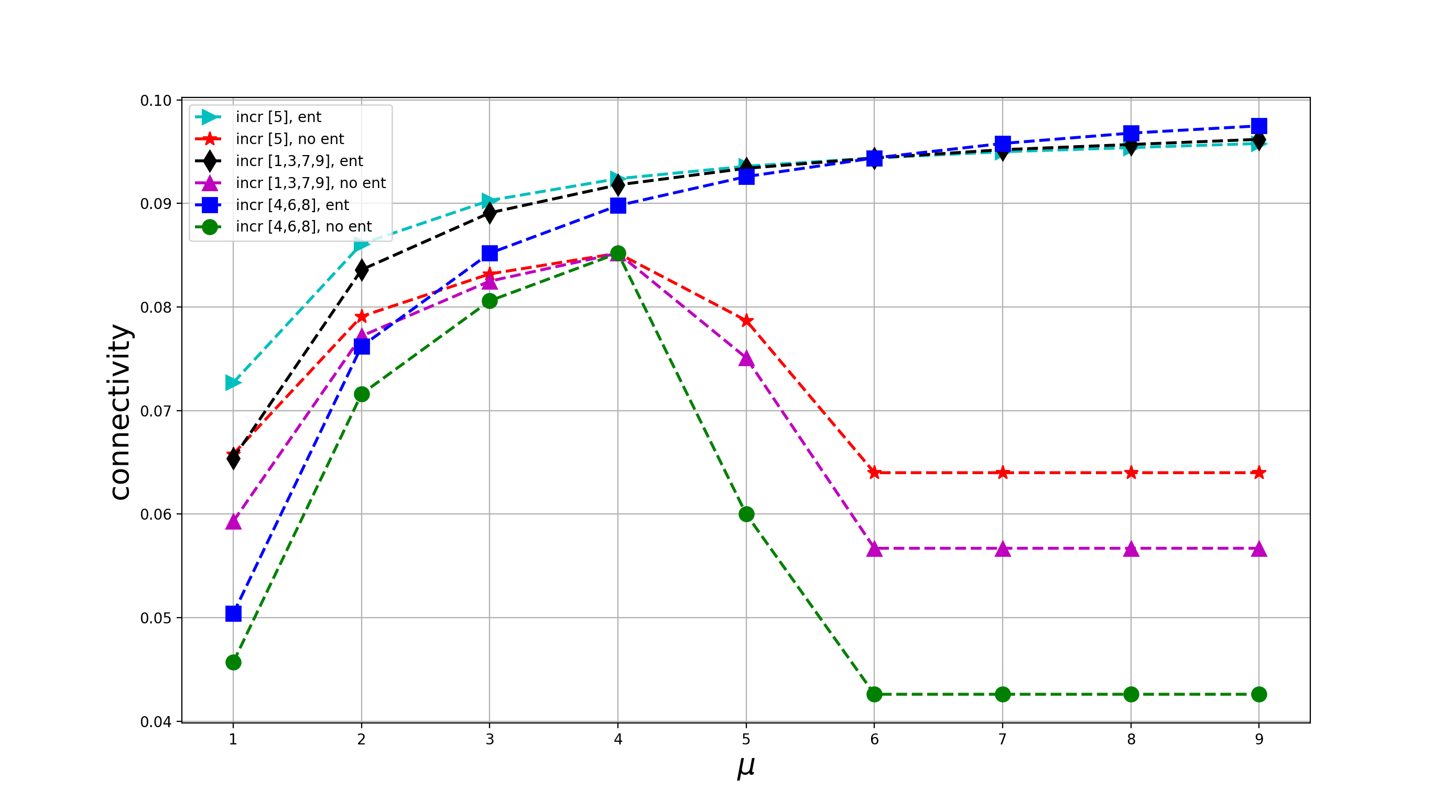}
\caption{Recover connectivity after removing node 2} \label{fig:recover}
\end{figure}

To relieve the harmfulness of removing node 2, Fig. \ref{fig:recover} shows one can increase the $\mu$ of node 5 before $\mu < 4$ for both entangle and non-entangled cases, and increase $\mu$ of nodes [4,6,8] for entangled case when $\mu \geqslant 4$. For non-entangled case, increasing node 5 saves more resources. 

\section {Conclusion}
Combined with algebraic connectivity, we have studied the connectivity of a quantum network based on quantum related physical parameters, such as the number of photons per pulse. The method we used can provide preliminary quantitative indication of which resources to use to strengthen the connectivity, and how much to it should be tuned.

\bibliographystyle{unsrt}  

\bibliography{bio}

\end{document}